\documentstyle[sprocl,epsfig]{article}

\def\bea{\begin{eqnarray}}
\def\eea{\end{eqnarray}}
\def\beq{\begin{equation}}
\def\eeq{\end{equation}}
\def\be{\begin{equation}}
\def\ee{\end{equation}}
\def\ben{\begin{equation}}
\def\een{\end{equation}}
\def\bc{\begin{center}}
\def\ec{\end{center}}
\def\dst{\displaystyle\strut}
\def\ov{\over\displaystyle\strut}

\def\t0{\tau_0}

\def\bea{\begin{eqnarray}}
\def\eea{\end{eqnarray}}
\def\l({\left(}
\def\r){\right)}

\def\bk{{\bf{k}}}
\def\bK{{\bf{K}}}
\def\bK{{\bf{K}}}
\def\bQ{{\bf{Q}}}

\def\dst{\displaystyle\phantom{|}}
\def\ov{\over\dst}

\def\xb{{\overline{x}}}
\def\tb{{\overline{t}}}
\def\rb{{\overline{r}}}
\def\nb{{\overline{n}}}
\def\etab{{\overline{\eta}}}
\def\taub{{\overline{\tau}}}

\def\t{{\tau}}

\def\ben{\begin{eqnarray}}
\def\enn{\end{eqnarray}}
\def\bea{\begin{eqnarray}}
\def\eea{\end{eqnarray}}
\def\be{\begin{equation}}
\def\ee{\end{equation}}
\def\ov{\over\displaystyle\strut}
\def\dst{\displaystyle\phantom{|}}
\def\l({\left(}
\def\r){\right)}

\def\bdk{{\bf \Delta k}}
\def\rl{R_l^2}
\def\ro{R_{o}^2}
\def\rs{R_{s}^2}
\def\rol{R_{ol}^2}

\def\BL{{Buda}{-}{Lund}~} 

\begin{document}

\title{INVARIANT BUDA - LUND PARTICLE INTERFEROMETRY}

\author{{T. CS\"ORG\H O$^1$ and B. L\"ORSTAD$^2$}  \\[1.5ex]}

\address{$^1$MTA KFKI RMKI,\\ 
	H-1525  Budapest 114, POB 49, Hungary \\
	E-mail: csorgo@sunserv.kfki.hu\\[2ex]
	$^2$Physics Department, Lund University,\\
	S -221 00 Lund, POB 118, Sweden\\
	E-mail: bengt@quark.lu.se} 
\maketitle
\abstracts{ 
The  invariant Buda-Lund parameterization of 
the two-particle Bose-Einstein correlation functions is presented, 
its derivation is summarized. 
In its  particular multi-variate Gaussian limiting case,
the invariant Buda - Lund parameterization is compared
to the Bertsch-Pratt and Yano-Koonin-Podgoretskii parameterizations,
advantages and shortcomings are discussed. 
The invariant Buda-Lund parameterizations are given
also for non-Gaussian multi-variate distributions,
including damped oscillations in the like-particle 
correlation function, that are similar to the oscillating
intensity correlations of binary stars in stellar interferometry.
A separation between the pion and the proton source is also
estimated in the Buda - Lund hydro framework, 
the result is utilized to extract the mean proper time of
particle emission with the help of fits to E877 data on non-identical particle
correlations by Miskowiec.
}

\section{Introduction}
Presently there is an increasing interest both 
in high energy heavy ion and in high energy particle physics
to describe in greater details the space-time picture of
particle emission with the help of particle interferometry:
the space-time geometry of the freeze-out hypersurface carries information
on the existence of a transient Quark-Gluon Plasma phase to be created
in heavy ion collisions. Overlapping regions of 
$(q\overline{q})+(q\overline{q})$ jets from a decaying $W^{+},W^{-}$
pair at LEP-II may result in unwanted systematic errors in precision
determination of $W$ mass in particle physics, and the magnitude of
this effect can be estimated with precision only if the space-time picture
of these reactions is reconstructed and the magnitude of Bose-Einstein
correlations between pions from different $W$-s is calculated correspondingly.

An invariant formulation of 
Bose-Einstein correlation functions was found by the Budapest-Lund
collaboration in refs.~\cite{3d,ismd95}.
This parameterization is referred as 
the Buda-Lund parameterization of the correlation function, or
BL in short.
Although the BL results are rather generic, they contain in 
particular limiting cases the {\it power-law}, the {\it exponential},
the {\it double-Gaussian}, the {\it Gaussian} or other parameterizations.
The BL parameterization of the correlation function yields  
not only a {\it boost-invariant}  description, but also a rather 
{\it simple} functional form.

We consider high energy heavy ion reactions or single jets
in high energy particle physics coming from a fragmentation
of an energetic leading quark. In these physical situations,
a dominant direction of expansion of the particle source
is identified. We denote space-time coordinates by
$x = (t,{\bf r}) = (t, r_x, r_y, r_z)$ and momentum variables
as $ k = (E, {\bf k} ) = (E, k_x, k_y, k_z)$. We choose the
direction labelled with subscript $z$ to coincide with the 
dominant direction of the expansion.

	The single-particle spectra and the two-particle
correlations are determined in the Wigner-function formalism.
As this formalism is fairly well-known now, we move 
the summary of the derivation  to section~{\ref{s:wig}},
and jump to the results immediately, as given in the next section. 
We then compare 
the BL form of the correlation function 
to the Bertsch - Pratt (BP) and the Yano- Koonin - Podgoretskii (YKP)
3-dimensional parameterizations
in a particular, Gaussian limiting case of the BL correlation function, 
as the BP and the YKP parameterizations are defined at the moment only 
for Gaussian correlation functions, as far as we know.
We also consider effective separation of sources in the 
BL hydro model, also described in ref.~\cite{3d}.
The effective separation of like-particle sources  
is shown to result in small, damped oscillations in the 
invariant longitudinal component of the two-pion intensity correlation
function. The effective separation of unlike-particle sources
in the BL hydro parameterization is larger.
Such a separation was recently determined experimentally by D. Miskowiec
et al in the analysis of E877 data on Au + Au collisions at BNL AGS,
based on their estimated $L = 10 $ fm/c separation 
scale and the BL hydro picture we estimate the mean freeze-out time
in Au + Au collisions.

\section{The \BL ~{parameterization for Bose-Einstein correlations}}
The two-particle correlation function is
defined as
\bea
 	C(\bk_1,\bk_2) & = & 
		{\dst N_2(\bk_1,\bk_2) \ov N_1(\bk_1) N_2(\bk_2) },
\eea
	a ratio of the two-particle invariant momentum distribution
	to the product of the single-particle invariant momentum
	distributions. Hence the correlation function is 
	invariant by definition, see ref.~\cite{voloshin} 
	for an exploration of its properties based on its 
	Lorentz invariance. 
	We assume, that with the help of certain experimental methods
	all non-Bose-Einstein related correlations can be removed
	from this function, and we focus on the Bose-Einstein
	correlations only, as these correlations carry information
	about the space-time distribution of particle emission.  
	The mean and the relative four-momentum 
	are introduced as
\bea
	K & = & (k_1 + k_2)/2, \\
	\Delta k  & = & k_1 - k_2.
\eea
	In the following, the relative momentum four-vector
	shall be denoted also as 
	$\Delta k  = Q = (Q_0, Q_x, Q_y, Q_z) = (Q_0, \bQ)$,
	and in general ${\bf a}$ shall denote the 
	vector part of a four-vector $a = (a_0, {\bf a})$.	

	A   simple invariant formulation of Bose-Einstein correlation
	functions was given for cylindrically symmetric,
	longitudinally expanding particle sources
	in refs.~\cite{3d,ismd95}. Such sources may be characterized 
	not only with longitudinal but also by transversal and
	temporal inhomogeneities.
	In a Gaussian approximation, the  Buda-Lund form of
	the Bose-Einstein correlation function reads as follows:
\be
	{C({\bf \Delta k},{\bf K}) = }
	{1}
	{+} \lambda \,
	{ \exp\left( - R^2_= Q^2_= - R^2_{\parallel} Q^2_{\parallel} - 
			R_{\perp}^2 Q_{\perp}^2 \right)} ,
\ee
	where the fit parameter	
	$\lambda$ measures a strength of the correlation function.
	The fit parameter  
	$R_=$ reads as $R$-timelike, and this variable
	measures a width of the proper-time distribution. 
	The fit parameter $R_{\parallel}$ reads as $R$-parallel, it
	measures an invariant
	length parallel to the direction of the expansion.
	The fit parameter $R_{\perp}$ reads as 
	$R$-perpedicular or $R$-perp. 
	For cylindrically symmetric sources, 
	$R_{\perp}$ measures a 
	transversal rms radius of the particle emitting source.

	The invariant time-like, longitudinal and transverse relative
	momenta are defined with the help of another, to this point
	suppressed  fit parameter, $\etab$,
	which characterizes the direction  of the center of particle emission,
	$\xb = (\tb, \rb_z)$ in the $(t,   r_z)$ longitudinal coordinate space.
	Such a direction can be characterized by a normalized four - vector 
	$\nb(\xb) = \xb / \taub$, where $\nb \cdot \nb = +1 $,
	and $\taub = \sqrt{\tb^2 - \rb_z^2}$ 
	is the mean longitudinal proper-time of particle emission.
	The such  a direction-pointing normal vector $\nb$ 
	can be parameterized as $\nb = (\cosh[\etab], 0, 0, \sinh[\etab])$,
	where $\etab = 0.5 \log\left[ (\tb+\rb_z)/(\tb - \rb_z)\right]$
	is the space-time rapidity of the center of particle
	emission, see Fig. 1. Space-time rapidity is a space-time coordinate,
	that transforms additively in case
	of longitudinal Lorentz-boosts, similary to the the rapidity
	variable in momentum space.
	A boost - invariant decomposition of the relative momenta can be defined,
	as follows. The invariant time-like, parallel and
	perp relative momentum components read as  
\bea
	Q_= & = &   
		Q_0 \cosh[\etab] - Q_z \sinh[\etab]\, \equiv  \,Q \cdot \nb 
		 \label{e:q=} \\
	Q_{\parallel} & = & 
		 Q_0 \sinh[\etab] - Q_z \cosh[\etab]  
		\, \equiv \,
		 Q \times \nb , \label{e:qpar} \\
	Q_{\perp} & = & \sqrt{ Q_{x}^2 + Q_{y}^2},   	\\
 	Q^2 & = & Q \cdot Q \, = \,  
	(Q\cdot \nb)^2 - (Q \times \nb)^2 - Q_{\perp}^2. \,   \label{e:qperp}
\enn
	In the above, 
	$a \cdot b = a^{\mu} b_{\mu} = a_0 b_0 - {\bf a} {\bf b}
	 = a_0 b_0 - a_x b_x - a_y b_y - a_z b_z$
	stands for the inner product of four-vectors. 
	As $\nb \cdot \nb = +1$, this   direction is time-like,
	hence $Q \cdot \nb =  Q_=  $ is an invariant time-like component of the 
	relative momentum.
	$Q_{\parallel}$ is an invariant relative momentum component
 	in the longitudinal direction (parallel to the beam
	or to the thrust axis),
	and $Q_{\perp}$ is the remaining perp or transverse 
	component of the relative momentum, also invariant for
	longitudinal boosts.	

	In short, we have the following {\it $5$ free parameters} for cylindrically
	symmetric, longitudinally expanding sources:
	$\lambda$, $R_=$, $R_{\parallel}$, $R_{\perp}$ and $\etab$.
	In principle, each of these parameters may depend on the mean momentum
	$\bK$. At any fixed value of the mean momentum $\bK$,
	the 5 free parameters of the invariant \BL correlation function
	can be fitted to data; alternalively they can be theoretically evaluated
	from an assumed shape of the emission function
	$S(x,k)$, for example see ref~\cite{3d}.
	In the \BL ~parameterization, the explicit mean momentum 
	dependence of the parameters can be written as follows: 
\be
	C({\bf \Delta k},{\bf K}) = 
	1 + \lambda({\bf K}) \exp\left( - R^2_=({\bf K}) Q^2_=({\bf K})
			- R^2_{\parallel}({\bf K}) 
			Q^2_{\parallel}({\bf K}) -
			R_{\perp}^2({\bf K}) Q_{\perp}^2 \right)   ,
\ee
	Note that the mean momentum dependence of the relative momentum
	components $Q_=(\bK) = Q_=(n(\bK))$ and 
	$Q_{\parallel}(\bK) = Q_{\parallel}(\nb(\bK))$ is induced by the
	mean momentum dependence of the direction pointing
	normal vector  $\nb(\bK)$, similarly to the dependence
	of the side and the out components of the relative momentum
	on the direction of the mean momentum in the Bertsch - Pratt
	parameterization.
	Physically, $R_=$, $R_{\parallel}$ and $R_\perp$ are
	longitudinally boost-invariant lengths
	of homogeneity~\cite{sinyukov} in the time-like, the
	longitudinal  and the transverse directions.
	Hence, $R_{\parallel}(\bK)$ is in general less 
	than the total longitudinal extension of the source. 
	Similar statements hold for the other,
	invariant lengths of homogeneity in the transverse
	and in the temporal directions, $R_{\perp}(\bK)$ and  $R_=(\bK)$.


\subsection{Symbolic notation for the side and out components}

Up to this point, we assumed a cylindrically symmetric source,
where the spatial information about the source distribution in $(r_x,r_y)$ 
was combined to a single  perp radius parameter $R_{\perp}$. 

In order to distinquish easily the zero-th component of the
relative momentum $Q_0$ from the out component of the
relative momentum $Q_o = Q_{out}$, $Q_0 \ne Q_o$,
we introduce the following  symbolic notation in transverse directions:
\bea
	Q_{side} & \equiv & Q_{\bf ..}\,\, , \\
	Q_{out}  & \equiv & Q_{\bf :}\,\, , \\
	Q_{\perp}^2 & = & Q_{side}^2 + Q_{out}^2 = Q_{\bf ..}^2 + Q_{\bf :}^2\,\,. 
\eea
The idea behind this notation is similar to that of symbolizing 
the perp direction by index ``$\perp$", standing for
a direction that is transverse to the longitudinal direction.
The longitudinal components of the relative momenta are symbolized by
index ``$\parallel$", i.e. two vertical parallel 
lines, imagining that the $z$ axis is in the plane of the paper,
pointing upwards, and the considered component is parallel to  
that direction. The invariant time-like components are  
indexed by ``$_=$", not to be confused by the equality sign $=$.
This symbol ``$_=$" is obtained 
by a 90$^o$ rotation of the  symbol ``$\parallel$". 
The remaining two orthogonal directions are the ``{\it side}" and the
 ``{\it out}", we symbolize them by parallel lines as well. But the
two possible orientations of parallel lines in the plane are used 
up by the symbols $_=$ and $\parallel$, hence these two lines are thought
to be orthogonal to the plane of the paper, thus they are symbolized
by two dots. The side component is described by putting the dots 
side-by-side, which yields index ``$_{..}$". The remaining out direction
is orthogonal to all the above directions, we rotate side by 90$^o$
to obtain the symbol ``$_:$".
This symbolic notation stands for 
the longitudinally boost-invariant decomposition of the relative momenta.
	Similarly, the side and the out radii can be denoted as
\bea
	R_{side} & = & R_{\bf ..}, \\
	R_{out} & = & R_{\bf :}.
\eea
	Cross-terms, if any, can be denoted by straightforward
	mixing of the symbols, e.g. a possible side-out cross-term
	may be denoted by ``$.:$", an out-long cross-term by ``$:|$" and
	a side-long cross-term by ``$..|$" etc.
	In a general Gaussian form, suitable for studying opacity
	effects, the \BL ~invariant
	BECF can be denoted as
\be
	C(Q,K) = 1 + \lambda_* \exp\left( - R_=^2 Q_=^2 -
			 R^2_{\parallel}  Q^2_{\parallel} -
			R_{..}^2 Q_{..}^2 - R_:^2 Q_:^2 \right)
\ee
	Note, that this equation is identical to eq. (44) of
	ref.~\cite{3d},  rewritten into the new, symbolic notation
	of the Lorentz-invariant directional decomposition. 

	The above equation may be relevant for a study of
	expanding shells as well as opacity effects as
	recently suggested by H. Heiselberg~\cite{henning}. 
	The lack of transparency in the source may result in an 
	effective source function, that looks like a crescent
	in the side-out reference frame~\cite{henning}. The 
	overall cylindrical symmetry of particle production is maintained
	for simultaneous rotations in the $x$ and the $k$ 
	space, but the emission function $S(x,K)$ at a given fixed direction
	of the mean transverse momentum $K_{\perp} = (0,K_x,K_y,0)$ 
	has different width in 
	the {\it out} direction and another, possibly larger 
	width in the {\it side} direction~\cite{3d}. 

\section{ General derivation of \BL ~shape -- not only for {Gaussians} 
	\label{s:wig}}

First we define the correlation function with the help of the
Wigner-function formalism, then we introduce the intercept 
parameter $\lambda_*$ in the core-halo picture.
Then we evaluate the correlation function
in terms of longitudinally  boost-invariant variables
and we end up with a general \BL ~form of the correlation function.

\subsection{Wigner Function Formalism}

The two-particle inclusive correlation function is defined and
approximately expressed in the Wigner function formalism as
\ben
C(\Delta k;K) & = & 
                {\dst N_2 ({\bf k}_1,{\bf k}_2)
                \ov N_1({\bf k}_1) \, N_1({\bf k}_2) }
                 \simeq
                1 + {\displaystyle\strut \mid \tilde S(\Delta k , K) \mid^2
                                \ov
                        \mid \tilde S(0,K)\mid^2 }.
	\label{e:cn1}
\enn
In the above line,
the Wigner-function formalism\cite{pratt_csorgo,zajc,chapman_heinz}
is utilized assuming fully chaotic (thermalized) particle emission.
The covariant Wigner-transform of the source density matrix,
 $S(x,k)$ is a quantum-mechanical  analogue of the classical
 probability that a boson is produced at a given
$ x^{\mu} = (t, {\bf r}) = (t,r_x,r_y,r_z)$
with $k^{\mu} = (E, {\bf k}) = (E, k_x, k_y, k_z)$.
The auxiliary quantity 
$$ \tilde S(\Delta k , K ) \, = \, 
\int d^4 x S(x,K) \exp(i \Delta k \cdot x )$$
appears in the definition of the BECF,
with $\Delta k  = k_1 - k_2$  and $K  = {(k_1 + k_2) / 2}$.
The single-  and two-particle inclusive momentum distributions (IMD-s)
 are defined in terms of the cross-section, they can be evaluated as
\ben
	N_1({\bf k})  =
	{\dst E \over \sigma_{t}} {\dst d \sigma \ov d{\bf k}}
		= \tilde S(\Delta k = 0, k)
	\quad & {\mbox{\rm and}} & \,\,\, 
	N_2({\bf k_1},{\bf k_2})  =  {\dst E_1 E_2 \over \sigma_{t}}
                 {\dst d \sigma \ov d{\bf k}_1 \, d{\bf k}_2},
\enn
where $\sigma_{t}$ is the total inelastic cross-section.
Note that in this work we utilize the absolute
normalization of the emission function:
$\int \!\!{\dst d^3{\bf k}\ov E}  d^4x  \, S(x,k) = \langle n \rangle $.
It has been shown recently in refs.~\cite{miskowiec,jzcst,cstjz,zhang_n}
that for explicitely symmetrized $n$-particle
system of bosons with variable number of bosons, 
the two-particle Bose-Einstein correlation function is 
properly defined by eq.~(\ref{e:cn1}).

\subsection{Effects from Large Halo of Long-Lived Resonances}

If the bosons originate from a core which is surrounded by
a halo of long-lived resonances, the IMD and the BECF can be
calculated in a straightforward manner. The detailed
description is given in refs.\cite{chalo,mhalo},
here we recapitulate only the basic idea along the lines of
ref.~\cite{chalo}.
{\it If} the emission function can be approximately divided into
two parts, representing the core and the halo,
        $ S(x;K)  =  S_{c}(x;K) + S_{h}(x;K) $
and {\it if} the halo is characterized
by large length-scales so that     $\tilde S_h(Q_{min};K) $
$ << \tilde S_c(Q_{min};K)$
at a finite experimental resolution of $Q_{min} \ge 10 $ MeV,
then 
\ben
        N_1({\bf k}) & = &  N_{1,c}({\bf k}) +  N_{1,h}({\bf k}), \\
        C(\Delta k;K) & = & 1 + \lambda_*
              {\dst \mid \tilde S_{c}(\Delta k, K) \mid^2 \ov
                        \mid \tilde S_{c}(0,K) \mid^2 },
			\label{e:chalo}
\enn
where $N_{1,i}({\bf k})$ stands for the IMD
of the halo or core for $i = h,c$ and
\ben
        \lambda_* & = & \lambda_*(K = k) = \left[{\dst N_{1,c}({\bf k}) \ov
                                 N_{1}({\bf k})} \right]^2.
\enn
The phenomenological $\lambda_*$ parameter has been
introduced to the literature by Deutschmann long time ago\cite{deutschmann}.
In the core/halo picture, this effective
intercept parameter $\lambda({\bf k})$ {\it can be interpreted}
as the {\it momentum dependent} square of the ratio of the IMD
of the core to the IMD of all particles emitted, assuming completely
incoherent emission from the source.

The validity of the core/halo picture to any given reaction is
not guaranteed, hence systematic checks of the applicability
of this simplifying picture has to be performed, similarly
to the ones done in ref.~\cite{dal}.

\subsection{Invariant, generic decomposition of the Bose-Einstein
	correlation function to a  Buda - Lund form }
For systems expanding relativistically in one direction ($r_z$) , 
it is advantageous to introduce the longitudinally boost invariant
variable $\tau$ and the space-time rapidity $\eta$ that transforms
additively under longitudinal boosts,
\bea
	\tau &	= & \sqrt{t^2 - r_z^2}, \\
	\eta & = & 0.5 \log\left[(t+r_z)/(t-r_z)\right] .
\eea
Similarly, in momentum space one introduces the transverse mass
$m_t$ and the rapidity $y$ as
\bea
	m_t & = & \sqrt{E^2 - p_z^2}, \\
	y & = &  0.5 \log\left[(E+p_z)/(E-p_z)\right] .
\eea
\begin{center}
\vspace{1cm}
\epsfig{file=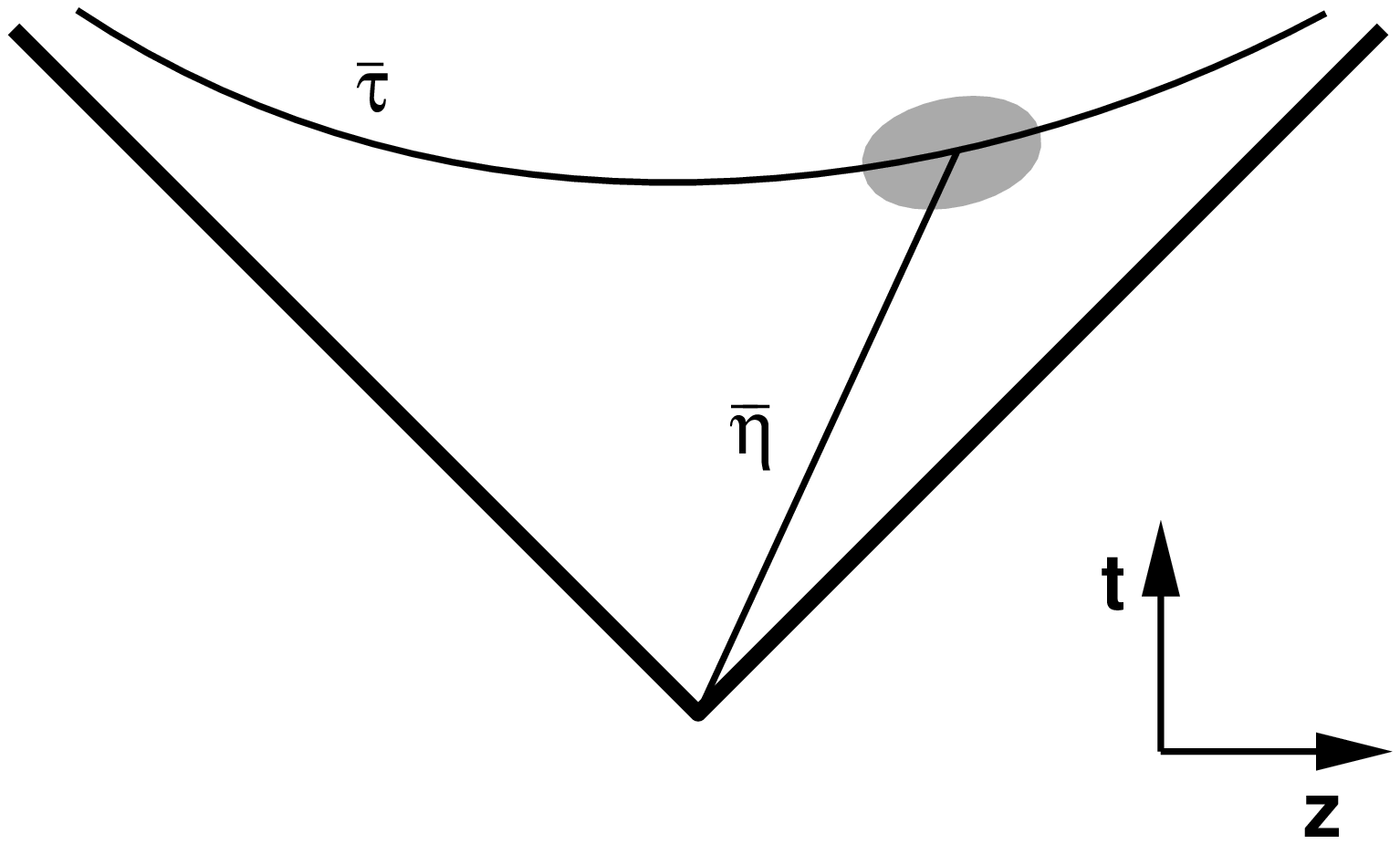,height=6cm,width=9cm,angle=0}\\
\begin{minipage}[t]{11cm}
	{\bf Fig. 1.} Space-time picture particle emission for
	a given fixed mean momentum of the pair. 
	The mean value of the proper-time  
	and the  space-time rapidity	distributions
	is denoted by $\taub$ and  $\etab$. 
	As the rapidity of the produced particles changes from the 
	target rapidity to the projectile rapidity 
	the $(\taub(y),\etab(y)$ variables scan the 
	surface of mean particle production in the $(t,r_z)$   plane.
\end{minipage}
\vspace{0.3cm}
\end{center}	

In order to obtain (at least approximately) a boost-invariant picture,
we characterize the source of particles in the boost invariant variables
$\tau$, $m_t$ and $\eta - y$. 
For systems that are only approximately boost-invariant, 
 the emission function may also 
depend on the  deviation from mid-rapidity $y_0$.
The scale on which the approximate boost-invariance breaks down
is denoted by $\Delta \eta$, a parameter is related to the 
	width of the rapidity distribution.

A simple generalization to non-Gaussian correlators
in the \BL~picture is obtained if we assume that  
the emission function factorizes
as a product of 
an effective proper-time distribution, a space-time rapidity
distribution and a transverse coordinate distribution
~\cite{lcms,3d} as 
\be
	S_c(x,K) d^4 x  =  H_*(\tau) G_*(\eta) I_*(r_x,r_y)\, d\tau \, \taub d\eta 
		dr_x dr_y .
\ee
	In the above, subscript $*$ stands for a dependence on the
	mean momentum, the mid-rapidity and the 
	scale of violation boost-invariance  $(K, y_0, \Delta\eta)$. 
	The function $H_*(\tau) $ stands for an effective proper-time 
	distribution (that includes an extra factor $\tau$ from the Jacobian
	$d^4x = d\tau\, \tau\, d\eta, dr_x dr_y$, in order to simplify
	the results).
	The effective (K dependent) space-time rapidity distribution
	is denoted by $G_*(\eta) $, while the  effective 
	transverse distribution is denoted by  $I_*(r_x,r_y) $ .
	In the above equation, the mean proper-time $\taub$ is
	factored out to keep the  dimensionless nature of the distribution 
	functions.
	Such a pattern of particle production is visualized
	on Fig. 1.
	
	In case of hydrodynamical models, as well as in case
	of a decaying Lund strings~\cite{lcms,ismd95},
	production of particles with a  given  momentum 
	rapidity $y$ is limited to a narrow region in 
	space-time around $\etab$ and $\taub$.
	If the sizes of such an effective source are sufficiently
	small (or with other words if the Bose-Einstein
	correlation function is sufficiently broad),
	the plane-waves that induce 
	the Fourier - transformation in the correlation function
	can be decomposed in the shaded region on Fig. 1
	as follows:
\be
	\exp[i \Delta E(t_1 - t_2) - i Q_z (r_{z,1} - r_{z,2}) ]  
	 \simeq  
	\exp[i Q_= (\tau_1 - \tau_2) - 
	i Q_{\parallel} \taub (\eta_1 - \eta_2) ], \label{e:blexp} \\
\ee
\bea
	\exp[ - i Q_x(r_{x,1} - r_{x,2}) - i Q_y (r_{y,1} - r_{u,2}) ]  
	 & = &  \nonumber \\
	&& \hspace{-2.5cm} =
	\exp[ - i Q_:(r_{:,1} - r_{:,2}) - i Q_{..} (r_{..,1} - r_{..,2}) ].
		\label{e:bltexp}
\eea
	With the help of this small source size (or large relative momentum)
	expansion, the two-particle Bose-Einstein correlation function
	can be written into the following \BL~form: 
\bea
	C(\Delta k, K) & = & 1 + \lambda_*(K) 
		{\dst |\tilde H_*(Q_=)  |^2 \ov |\tilde H_*(0) |^2} \,
		{\dst |\tilde G_*(Q_{\parallel})  |^2 \ov |\tilde G_*(0) |^2}\,
		{\dst |\tilde I_*(Q_{:},\, Q_{..})|^2 \ov 
		|\tilde I_*(0;0) |^2} .
	\label{e:blf}
\eea
	This is the factorized Buda-Lund invariant 
	decomposition of the BE correlation function;
	the resulting Fourier-transformed proper-time	,
	space-time rapidity and transversal coordinate
	distributions can be of power-law, exponential,
	Gaussian or other types, corresponding to the
	underlying space-time structure of the particle
	emitting source: 
\bea
	\tilde H_*(Q_=)  & = & 
		\int_0^\infty d\tau \exp(i Q_= \tau) H_*(\tau),
			\label{e:htild} \\
	\tilde G_*(Q_\parallel)  & = & \int_{-\infty}^{\infty} d\eta 
		\exp(- i Q_\parallel \taub \eta ) G_*(\eta),
			\label{e:gtild}\\
	\tilde I_*(Q_:,Q_{..}) 
			 & = & \int_{-\infty}^{\infty} 
			d\tau \exp(- i Q_: r_: - i Q_{..} r_{..}) 
			I_*(r_:,r_{..}) . \label{e:itild}
\eea
	What are the fit parameters in eq.~(\ref{e:blf})?
	It is clear that apart from the shape parameters of the 
	proper-time, space-time rapidity and the transverse
	distribution of the production points two additional
	space-time parameters enter the fit: the mean
	proper-time of the particle production, $\taub$ in eq.~(\ref{e:gtild}) 
	and
	the angular direction $\etab$ that enters the definition
	of $Q_=$ and $Q_{\parallel}$ in eqs.~(\ref{e:q=}),\ref{e:qpar}). 
	In turn, the parameters $(\taub,\etab)$
	are measurable from the detailed analysis 
	of the multi-dimensional Bose-Einstein correlation
	functions for any value of the rapidity 
	and $m_t$ of the particle pair. The total longitudinal region
	corresponds to the region in $(t,r_z)$ where
	particles of arbitrary momenta are emitted from. 
	This region can be reconstructed in the \BL formalism, 
	by determining $(\taub,\etab) $ and the widths of 
	$H_*(\tau)$ and $G_*(\eta)$ at various fixed 
	values of the momentum of the particles,
	reproducing the shaded region of Fig. 1 for
	each fixed value of the mean momentum of the pair.
	Such shaded regions are the same as local maps in
	cartography.
	If the momentum distribution of the produced particles 
	is integrated over, the overlapping shaded regions
	are combined to a global picture of particle emission
	in space-time, similarly the way how local maps  can be combined
	to an atlas in cartography by gradually displacing the centers
	of the local maps but keeping an overlap region between the 
	neighbouring local pictures.

\begin{center}
\epsfig{file=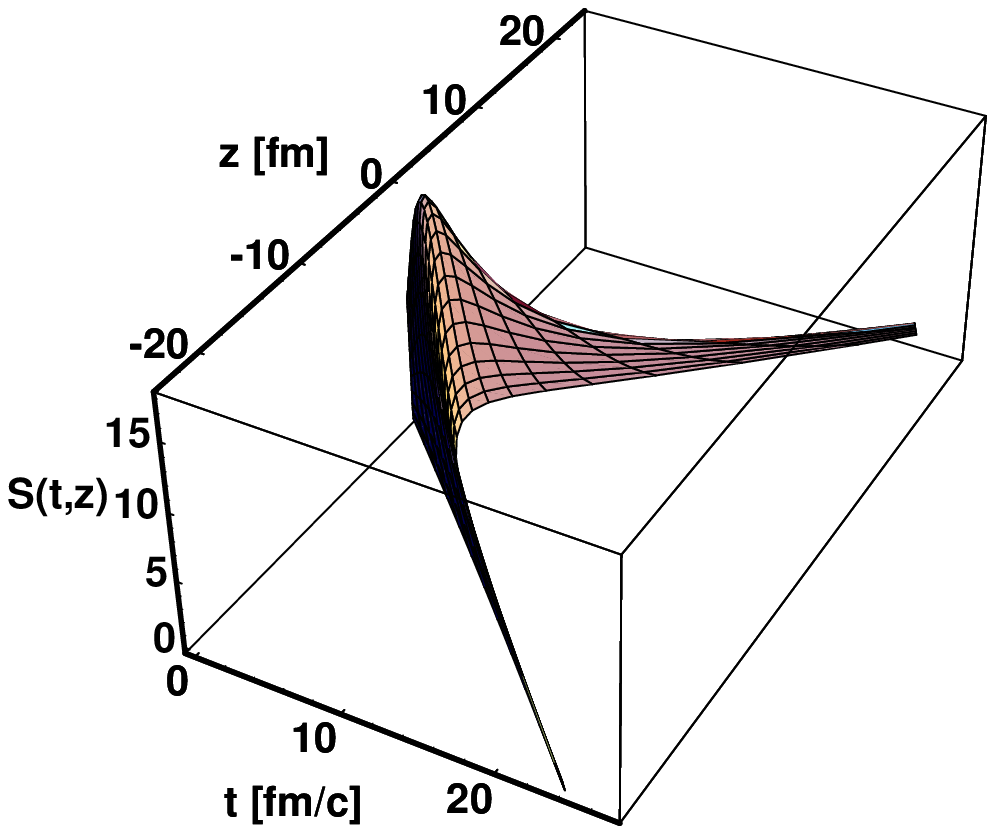,width=9cm,angle=0}\\
\begin{minipage}[b]{11cm}
	{\bf Fig. 2.} Space-time picture of particle emission 
	as reconstructed from a combined analysis~\cite{nr,1d,3d} of 
	particle correlations and spectra in $h + p$ reactions at CERN SPS 
	in $(t,r_z)$ plane. The momentum of the emitted
	particles is integrated over, from ref.~\cite{hakob-cf98}.
\end{minipage}
\vspace{0.3cm}
\end{center}	
	This programme has been carried out first by the NA22  collaboration
	in a Gaussian approximation, keeping the means and the variances
	only of the proper-time distribution, the space-time rapidity
	distribution and the transverse distribution coordinates
	of particle production. A locally thermalized, longitudinally
	expanding  medium 
	( possibly corresponding to the vacuum filled by the sea of 
	virtual quarks, anti-quarks and gluons) was found to be 
	in a good agreement with the NA22 particle spectra and correlations,
	despite the fact that the mean multiplicity of the 
	produced charged particles is only $\langle n \rangle = 8$.
	See R. Hakobyan's contribution to this conference proceedings
	for the reconstructed space-time picture as well as for
	greater details.

	A similar reconstruction of the longitudinal space-time
	structure of particle production in 200 AGeV $S + Pb$ reactions
 	is reported in the contribution of Ster to this volume.

	For even further general possibilities about the 
	structure of particle emission, see refs.~\cite{3d,lcms,ismd95}.

\section{Earlier, Gaussian parameterizations of BE Correlations}

	We briefly summarize here the Bersch-Pratt and the
	Yano-Koonin parameterization of the Bose - Einstein correlation
	functions, to point out some of their advantages as well as draw-backs 
	and to form a basis for comparision.

\subsection{The Bertsch-Pratt parameterization}
	The Bertsch-Pratt (BP) parameterization of 
	Bose-Einstein correlation functions is one of the oldest, 
	widely used parameterization called also as side-out-longitudinal
	decomposition~\cite{bertsch,pratt} of the correlation function.
 
	This directional decomposition was originally devised
	for extracting the contribution of the long duration
	of particles from a decaying Quark-Gluon Plasma,
	as happens in the mixture of a hadronic and a QGP phase
	if the rehadronization phase transition is a strong first order 
	transition. 
	
	The BP parameterization in a compact form reads as
\ben
    C({\bf \Delta k},{\bf K})   & = & 
	1 +  \lambda \exp\left[ - \rs Q_{s}^2 - \ro Q_{o}^2 
	- \rl Q_l^2 - 2 \rol Q_l Q_{o} \right].
\enn
	Here index $o$ stands for {\it out} (and not the temporal
	direction), $s$ for {\it side} and $l$ for {\it longitudinal}.
	In a more detailed form, the mean momentum dependence of 
	the various components are explicitly shown:
\ben
    C({\bf \Delta k},{\bf K})   & = & 
	1 +  
	\lambda({\bf K}) \exp\left[ - \rs({\bf K}) Q_{s}^2({\bf K})
        - \ro({\bf K}) Q_{o}^2({\bf K}) \right. \nonumber  \\
	\null & \null & \hspace{3cm} \left. 
	- \rl({\bf K}) Q_l^2 - 2 \rol({\bf K}) Q_l Q_{o}({\bf K}) \right]\, ,
\enn
	where the mean and the relative momenta are defined as
\ben
	{\bf K} & = & 0.5 ({\bf k}_1 + {\bf k}_2 ) ,\\
	{\bf \Delta k} & = & {\bf k}_1 - {\bf k}_2 , \\
	Q_l & = & k_{z,1} - k_{z,2} ,\\
	Q_{o} & = & Q_{o}({\bf K}) 
		\, = \, {\bf \Delta K} \cdot {\bf K} / | {\bf K} | , \\ 
	Q_{s} & = & Q_{s}({\bf K}) \, = \,
		|{\bf \Delta k } \times {\bf K} | / | {\bf K} | . 
\enn
	It is emphasized that the BP radius parameters
	are  measuring lengths of homogeneity, and in general
	characterize that region of space-time that emits particles
	with a given mean momentum  ${\bf  K}$. 
	Not only the radius parameters but also the decomposition of the relative momentum to the side
	and the out components depends on the mean momentum ${\bf K}$. 

	In arbitrary frame, Gaussian radius parameters 
	can be defined, and sometimes they are also referred to
	BP radii, when the spatial components of the relative momentum
	vector are taken as independent variables.
	It should be emphasized, that the BP radii 
	that contain a {\it well-defined mixture }
	of the longitudinal, temporal and transverse
	invariant radii. 
	However, the BP radius  parameters themselves are not invariant,
	they depend on the frame where they are evaluated~\cite{3d},
	reflecting space-time variances~\cite{uli_s,uli_l} 
	of the {\it core}~\cite{dal} of the particle emission.
\bea
	C^{c/h}(\bk,\bdk ) & = & 1 + \lambda_*({\bf K}) \,
	\exp\left(- R^2_{i,j}({\bf K}) {\bf \Delta k}_i {\bf \Delta k}_j \right), \\
	\lambda_*({\bf K}) & = & [ N_{\bf c}({\bf K}) / N({\bf K}) ]^2, \\
	R_{i,j}^2({\bf K}) & = &
		 \langle (x_i - \beta_i t) (x_j - \beta_j t) \rangle_{\bf c}
		 - \langle (x_i - \beta_i t)\rangle_{\bf c}
		   \langle (x_j - \beta_j t) \rangle_{\bf c}, \\
	\langle f(x,{\bf k}) \rangle_{\bf c} & = & 
		\int d^4 x f(x,{\bf k})  S_{\bf c}(x,{\bf k})  /
		\int d^4 x  S_{\bf c}(x,{\bf k}),
\eea
	where 
	$S_{\bf c}(x,{\bf k})$ is the emission function 
	that characterizes the central core. Note, that the tails
	of the emission function are dominated by the halo of long-lived
	resonancess $S_h(x,{\bf k})$ and even a small admixture
	of e.g. $\eta$ and $\eta'$ mesons increases drastically the
	space-time variances of particle production, 
	and makes the interpretation of the BP radii in terms
	of the total emission function $S = S_c + S_h$ unreliable,
	as pointed out in ref.~\cite{dal}.

	It was shown already in ref.~\cite{lcms} that the duration of the
	particle emission contributes predominantly  to the out direction,
	if the Longitudinal Center of Mass System
	(LCMS, ref.~\cite{lcms}) is selected for the determination of the radius parameters.
	In LCMS, the mean momentum of the pair has no longitudinal
	component. In this frame, the BP radii have a particularly simple
	form, if the coupling between the $r_x$ and the $t$ coordinates
	is also negligible, $\langle \tilde r_x \tilde t \rangle = 
	\langle \tilde r_x \rangle \langle \tilde t \rangle$: 
\bea
	R_{out}^2({\bf K}) & = &
		 \langle \tilde{r}_x^2 \rangle_{\bf c} - 
		 \beta_t^2 \langle \tilde{t}^2 \rangle_{\bf c} \\
	R_{side}^2({\bf K}) & = &
		 \langle \tilde{r}_y^2 \rangle_{\bf c} \\
	R_{long}^2({\bf K}) & = &
		 \langle \tilde{r}_z^2 \rangle_{\bf c} \\
	R_{out,long}^2({\bf K}) & = &
		 \langle \tilde{r}_z 
		(\tilde{r}_x - \beta_t \tilde{t}) \rangle_{\bf c} ,
\eea
	where $\tilde x = x - \langle x \rangle $.
	From the above, the advantage of  the LCMS frame is clear: 
	In this LCMS frame, information on the temporal scale
	couples {\it only} to the out direction and it enters both the
	out radius component and the out-long cross-term.
	Note that for cylindrically symmetric sources
	other possible cross-terms, e.g. the side-out or the side-long
	cross terms were shown to vanish in a Gaussian approximation, 
	due to symmetry reasons~\cite{uli_s,uli_l}.

	An advantage of the BP parameterization is that
	there are no kinematic constraints between the side, out 
	and long components of the relative momenta,  
	hence the BP radii are not too difficult to determine experimentally.
	 
\subsection{The Yano-Koonin-Podgoretskii parameterization}

	A covariant parameterization has been worked out for 
	non-expanding sources by Yano, Koonin and Podgoretskii (YKP)
	~\cite{ykp1,ykp2}.
	This parameterization was recently applied to expanding
	sources by the Regensburg group~\cite{ykpr1,ykpr2},
	by allowing the YKP radius and velocity parameters be momentum
	dependent. This parameterization reads as
\bea
    C({\bf Q},{\bf K}) \!  & = & \!
	1 + \exp\left[ - R_{\perp}^2 Q_{\perp}^2 - 
		    R_{\parallel}^2  (Q_l^2 - Q_0^2 )
	- \left(R_0^2 + R_{\parallel}^2 \right)
	\left( Q \cdot U \right)^2 \right] , 
\eea
	(Note that in YKP index $_0$ refers to the time-like components).
	When the momentum dependence of the YKP radii is
	explicitly shown, this reads as
\bea 
    C({\bf Q},{\bf K})  & = & 
	1 + \exp\left[ - R_{\perp}^2({\bf K}) Q_{\perp}^2 - 
		    R_{\parallel}^2({\bf K})  (Q_z^2 - Q_0^2 )
	\right. \nonumber\\ 
	\null & \null  &  \hspace{2cm}
	\left. - \left(R_0^2({\bf K}) + R_{\parallel}^2({\bf K}) \right)
	\left( Q \cdot U({\bf K}) \right)^2 \right] ,
\enn	
	where the fit parameter $U({\bf K})$ is interpreted~\cite{ykpr1,ykpr2} 
	as a {\it four-velocity} of a fluid-element~\cite{seyboth-cf98}. 
	This YKP parameterization
	is introduced to create a diagonal Gaussian expression 
	in the ``rest frame of the fluid-element''. 

	This form has an advantage as compared to the BP parameterization,
	namely that the three extracted YKP radius parameters,
	$R_{\perp}$, $R_{\parallel}$ and $R_0$ are invariant, 
	independent of the frame where the analysis is performed,
	while $U^{\mu}$ transforms as a four-vector. 
	
	The practical price one has to pay for this advantage is that the 
	kinematic region in the $Q_0$, $Q_l$, $Q_{\perp}$
	space, where the parameters can be fitted
	may become rather small.  This follows from
	the unequalities $Q_{inv}^2 \ge 0$ and $Q^2_0 \ge 0$,
	which yields
\be
	0 \le  Q_0^2  \le Q_z^2 +	 Q_{\perp}^2,
\ee
	and the narrowing of the regions in  $Q^2_0 - Q_z^2$ 
	with decreasing $Q_\perp$ 
	makes  the experimental determination
	of the YKP parameters difficult, especially when the analysis is
	performed far away from the LCMS rapidities [or more precisely
	from the frame where $U^{\mu} = (1,0,0,0)$\, ].	Hence in 
	practice the YKP parameters can be well determined in the LCMS
	frame, where the longitudinal component of the $U$ is 
	generally small. But in the LCMS, the interpretation of
	the BP radii is also simple, similarly  
	to that of the YKP radii.

	Theoretical problems with  the YKP parameterization 
	are explained below. 
	{\it a)} The YKP radii contain components proportional to 
	${1 \over \beta_t}$, that lead to divergent terms 
	for particles with very low $p_t$ ~\cite{ykpr1,ykpr2}.
	{\it b)} The YKP radii are not even defined for all Gaussian
	sources~\cite{ykpr1,ykpr2}. Especially, for opaque
	sources or for expaning shells with 
	$\langle \tilde r_x^2\rangle < \langle \tilde r_y^2 \rangle$
	 the algebraic relations defining the YKP ``velocity" parameter 
	become ill-defined and result in imaginary values of the 
	YKP ``velocity",~\cite{ykpr1,ykpr2}.
	{\it c)} The YKP ``velocity" is defined in terms of
	space-time variances at fixed mean momentum of the particle pairs
	~\cite{ykpr1,ykpr2}.  Thus, for expanding systems, 
	the proper interpretation
	of the   parameter $U$  is {\it not} a flow velocity of 
	a source element, as thought before, but a combination
	of space-time variances of the source at a 
	fixed mean momentum ${\bf K}$ .
	(Note, however, that for static, 
	non-expanding sources the interpretation of $U^{\mu}$ as
	the velocity of a Gaussian source can be preserved
	corresponding to refs.~\cite{ykp1,ykp2}). 

	In kinetic theory that provides the basis for hydrodynamics, 
	the flow velocity can be locally defined as 
	a weighted average of particle {\it momenta}, 
	all particles being in the same cell in coordinate space. 
	The local flow velocity $u^{\mu}(x)$ hence becomes 
	a function of the position $x$ 
	but the momentum of the particles was averaged out,
	hence $u^\mu(x)$ is formally independent of the momentum.
	The four-current is defined~\cite{degroot} as an average over the
	local momentum distribution
\begin{equation}
	j^{\mu}(x) = \int {d^3k \over k^0} k^{\mu} f(x,k)
\end{equation}
	where $f(x,k)$ stands for the phase-space distribution function.
	The local flow velocity can be defined  with this current as 
	the unit vector proportional to the current:
\begin{eqnarray}
	u^{\mu}(x) & = & j^{\mu}(x) / \sqrt{j^{\mu}(x) j_{\mu}(x)}.
\end{eqnarray}
	
 	On the other hand, the YKP parameter $U$ 
	corresponds to a weighted average of particle {\it coordinates},
	all particles being characterized with the momentum ${\bf K}$.
	Hence in general $U(\bK) \ne u(x)$.
	 The local velocity of particles at a fixed momentum is independent
	of the density distribution of particle production
	in   coordinate-space.
	In fact, the only four-velocity that can be uniquely assigned
	to a set of particles each with momentum ${\bf K}$
	is simply $u^{\mu}_{\bf K} = (E_{\bf K}/m,{\bf K}/m)$, where $m$
	stands for the mass of the particles.
	Hence the YKP parameter $U({\bf K})$
	should not be interpreted as a mean {\it flow} velocity
	of a fluid element that emits particles with momentum ${\bf K}$,
	at least not in the well-defined sense of kinetic theory.   

\subsection{ Comments on the hydro model parameterization
}
The \BL hydro parameterization is recapitulated in
the contributions of Ster and Hakobyan
~\cite{ster-cf98,hakob-cf98}. The main idea behind the
\BL parameterization is to characterize
the hydro fields (temperature profile, transversal flow, density
distribution) with { their means and variances} only. 
Similar models are studied by the Regensburg~\cite{uli_s,uli_l,ykpr2} as well 
as the Kiev groups~\cite{yuri-hydro}.
	
	Note that ${\nb^\mu({\bf K})}$ normal-vector rests in
	the $(t,r_z)$ plane for the Buda-Lund parameterization.
	On the other hand,
	the	$U^{\mu}({\bf K})$ YKP velocity was thought to
	be some sort of flow velocity, that characterizes
	 some local momentum distribution.  Hence,
	$\nb^{\mu}$ and $U({\bf K})$ are defined in
	{\it different spaces}: in the coordinate space and in the momentum
	space, respectively. In case of the Buda-Lund form,
	the coordinate-space interpretation of $\nb$ is 
	needed to obtain the expansion of the plane-wave
	$\exp(i \Delta k \cdot \Delta x)$ in eq.~(\ref{e:blexp}),
	which is essential in expressing the correlation function
	in terms of Fourier-transformed proper-time distributions
	and space-time rapidity distributions in eq.~(\ref{e:blf}).
	If the space-time interpretation of the \BL direction
	$\nb({\bf K})$ is lost, it becomes impossible to
	reconstruct the space-time picture of particle emission
	for systems with strong longitudinal expansion. 
	That could be the reason why such  
	a reconstruction was not yet achieved by the NA49 experiment,
	 that applied the YKP parameterization. 
	However, the space-time picture of
	longitudinally expanding particle emitting sources
	was reconstructed from the detailed fitting to NA22  and NA44
	 particle correlations   and spectra with the help of the
	\BL ~parameterization, properly preserving the interpetation of
	of $\nb(\bK)$ as a spatial direction. 
	See Fig. 2, and the contributions of Hakobyan, Ster and Seyboth to this
	conference proceedings~\cite{hakob-cf98,ster-cf98,seyboth-cf98}.

\section{Separation of particle sources in \BL type hydro models}

We discuss some results relating the separation of effective sources
	for identical and non-identical particles, even if both kind
	of particles appear from the same system that is assumed to 
	be in local thermal equilibrium.
\subsection{Separation of effective sources for non-identical particles}
Recently, Lednicky and collaborators
suggested to study  non-identical particle-correlations,
to learn which particles are emitted earlier and which
particles were emitted later~\cite{lednicky}.
The analysis of E877 data resulted in an
effective separation of pion and proton sources at forward rapidities
at the AGS~\cite{miskowiec-el}.

Such separation of pion and proton source
in space-time occurs as a natural result in 
the \BL hydro models and its various re-incarnations and 
	improved modifications, because heavier particles
	are more frozen to the flow than the lighter ones.

In the \BL parameterization, we have a $K$ dependent normalvector,
pointing towards the center of the particle production.
 
In the  LCMS, ref~\cite{lcms} ,
we find
\bea
	n & = & n({\overline x}({\bf K}) ) = 
		(\cosh[\etab], 0, 0, \sinh[\etab] ) \\ 
	\etab & = & 
	\etab({\bf K}) \, = \, 
		 \frac{y_0 - y}{ 1 + \Delta\eta^2 m_t / T_0 } ,
\enn
	where $ T_0 $ is the central freeze-out temperature at the
	mean freeze-out time, and $\Delta\eta$ characterizes the
	finite longitudinal size of the expanding hydro source in
	space-time rapidities. 
	
	For non-identical particle correlations~\cite{lednicky}, e.g. $\pi$ and $p$, the 
	velocities of the particles must be similar.

If $v_{\pi} = v_p$, then we have $m_t^{\pi} << m_t^p$.
	For pions and protons with the {same} velocity, 
	using $m_{\pi} = 140$ MeV, $m_p = 1 $ GeV, $T = 140$ MeV and
	$\Delta\eta^2 = 2$ the \BL  model yields:

\be
	\etab^{p} - \etab^{\pi} \simeq {\dst y - y_0 \ov 3 } 
\ee 
	where $y_0$ is the mid-rapidity, $y$ is the rapidity
	of the pion and the proton, and 
	$3$ is the numerically estimated coefficient.
	
	The spatial separation between protons and pions is
	about $L \simeq 10$ fm/c at
	$y - y_0 \simeq  2$~\cite{miskowiec-el}. 
	This can be used to estimate the 
	mean freeze-out proper-time of particle production
	as { $\tau_s = L / (\eta_s^p - \eta_s^{\pi} ) \simeq 15$ fm/c.}

\subsection{Example for a non-Gaussian \BL correlation function:
	separation of the effective source for identical pions}

	In the \BL type hydro models, the emission function
	$S(x,k)$ is written as a product of the local phase-space
	density $f(x,k) $ and a Cooper-Frye pre-factor 
	$d^4 \Sigma^{\mu}(x) k_\mu$ that
	yields the flux of particles through the freeze-out 
	hypersurface, or through a distribution of freeze-out 
	hypersurfaces~\cite{3d}:
\bea
	S(x,k) d^4x & = & f(x,k) \, d^4 \Sigma \cdot k \\
	f(x,k)	& = & {\dst g \ov (2 \pi)^3 } 
		{\dst 1 \ov \exp\left[ {k \cdot u(x) - \mu(x)\ov T(x) }\right] + s}   \\
	d^4 \Sigma(x)\cdot k & = & m_t \cosh[\eta - y] H_*(\tau) 
		\, d\tau\,	\taub d\eta \, dr_x \, dr_y 
\eea
	where $g $ stands for the degeneracy factor,
	and $s = -1$, $0$, $+1$ for Bose, Boltzmann or Fermi statistics,
	and an approximate boost-invariant shape of the freeze-out
	hypersurface distribution is assumed.
	Using the exponential form of the $\cosh[\eta -y]$ factor,
	the effective emission function $S_*(x,k) = 
	{\tau\over \taub} S(x,k)$ can be written as a sum of two
	components:
\bea
	S(x,k) & = & 0.5 [ S_+(x,k) + S_-(x,k) ] ,\\
	S_+(x,k) & = & m_t \exp[+\eta - y] H_*(\tau) f(x,k) ,\\
	S_-(x,k) & = & m_t \exp[-\eta + y] H_*(\tau) f(x,k) .
\eea 
	These effective emission function components are subject to
	Fourier - transformation in the \BL approach.
	In an improved saddle-point approximation, the two components
	$S_+(x,k)$ and $S_-(x,k)$ can be Fourier - transformed 
	independently, finding the separate
	maxima (saddle point) $\xb_+$ and $\xb_-$ of
	$S_+(x,k)$ and $S_-(x,k)$, and repeating the saddle-point
	calculation for the two components separately. As a result, one gains
	the following non - Gaussian correlation function
	from the Buda-Lund hydro model specified in ref.~\cite{3d}
\bea
	C(Q_=,Q_{\parallel},Q_{..},Q_{:}) & = &
	1 + 
	 \lambda_* \, \Omega(Q_{\parallel}) \, 
		\exp( - Q_{\parallel}^2 R_{\parallel}^2
		 - Q_{=}^2 R_{=}^2
		 - Q_{\perp}^2 R_{\perp}^2 ), \label{e:dosc}\\
	\Omega(Q_{\parallel}) & = &  [ \cos^2( Q_{\parallel} R_{\parallel} 
				\Delta\overline{\eta} )
		+ \sin^2(Q_{\parallel} R_{\parallel} \Delta\overline{\eta}) 
		  \tanh^2( \overline{\eta}) ],  \\
	{1 \over \Delta\overline{\eta}^2} & = &  
	{1 \over \Delta\eta^2} + {m_t \over T_0} \cosh[\etab].
\eea
	This result goes beyond the single Gaussian version of the saddle-point
	calculations of ref.~\cite{uli_s,uli_l}. 
	This results goes also beyond the results obtainable in
	the YKP  or the BP parameterizations. In principle,
	the improved saddle-point calculation gives more accurate 
	analytic results than the numerical evaluation of space-time variances,
	as it keeps more information on the shape of the correlation
	function.

	Although the above result is non-Gaussian,  because the factor
	$\Omega(Q_{\parallel})$ results in oscillations of the correlator, 
	the result is still explicitely boost-invariant.
		
	Note that the oscillations are typically small and the
	Gaussian remains a good approximation to eq.~(\ref{e:dosc}),
	but with modified radius parameters. 

	The oscillations are due to a possible separation
	of the pion source to two components, due to an
	identical splitting of the Cooper-Frye
	pre-factor or the flux term in the emission function.
	We obtain two separate effective sources that
	create oscillations in the intensity correlation function,
	similarly to the oscillations in the intensity correlations
	of photons from binary stars in stellar astronomy.
	However, these oscillations in high energy physics are
	much smaller, as the effective separation between the 
	particle sources for identical pions, $\etab_+ - \etab_-$ 
	smaller, than $\Delta\etab$, the width of the 
	$G_{*+}(\eta_+)$
	and the  $G_{*-}(\eta_-)$ distributions. In stellar astronomy,
	the separation between the binary stars is much larger than the
	diameter of the stars, hence the intensity oscillations in the
	two-photon correlation function are stronger then in the \BL type
	hydro models.

\section{Highlights: }
	The invariant Buda-Lund notation is introduced 
	to describe Bose-Einstein correlation functions. This invariant
	notation scheme yields simple expressions not only for
	Gaussian but also for non-Gaussian expanding sources as well.
	It seems that the Buda-Lund form is the simplest and the most 
	compact characterization
	of the two-particle Bose-Einstein correlation functions
	for relativistic, expaning systems. 

	With the help of the \BL formulation, and a combined
	analysis of particle correlations and spectra~\cite{3d} ,
	the space-time picture of the particle production 
	in the longitudinal $(t,r_z )$ plane can be reconstructed. 
	Examples for such a reconstruction are shown 
	in refs.~\cite{ster-cf98,hakob-cf98}.

	We pointed out how to estimate the mean freeze-out time	
	using non-identical particle correlations data, the Buda-Lund
	hydro model and the natural separation of forward moving
	protons and pions in Buda-Lund type hydro models.

	Finally we have shown how \BL type hydro models can be
	rewritten to an effective, two-components sources,
	by splitting the flux terms. We found small damped oscillations in
	the intensity correlation function, reminiscent to the
	intensity correlations of photons from binary stars.

\section*{Acknowledgments}
I would like to thank to B. L\"orstad 
for a stimulating and frutiful collaboration,
and for his hospitality during my visits to Lund University.
I thank also O. Smirnova for her intriguing suggestions related to the
Buda-Lund notation.
This work was supported in part by 
the US - Hungarian  Joint Fund MAKA 652/1998, 
by the National Scientific Research Fund (OTKA, Hungary) Grant T026435,
and by NWO (Netherland) - OTKA grant N25487.

\end{document}